\documentclass[twocolumn, 10pt, prl, aps, showpacs, reprint, amsmath, amssymb, superscriptaddress, nofootinbib]{revtex4-1}

\usepackage[pdftex]{graphicx}

\begin{document}

\newcommand{\kt}{k_\parallel}
\newcommand{\lzt}{q_z}
\newcommand{\atled}{\bm{\nabla}}
\newcommand{\dx}{\frac{\partial}{\partial_x}}
\newcommand{\dy}{\frac{\partial}{\partial_y}}
\newcommand{\dz}{\frac{\partial}{\partial_z}}
\newcommand{\dt}{\frac{\partial}{\partial_t}}
\newcommand{\sqrdt}{\frac{\partial^2}{\partial_t^2}}
\newcommand{\pbyp}[2]{\frac{\partial #1}{\partial #2}}
\newcommand{\dbyd}[2]{\frac{d #1}{d #2}}
\newcommand{\ex}{\vec{e}_x}
\newcommand{\ey}{\vec{e}_y}
\newcommand{\ez}{\vec{e}_z}
\newcommand{\besselj}[2]{\mathrm{J}_{#1}(#2)}
\newcommand{\besseljp}[2]{\mathrm{J'}_{#1}(#2)}
\newcommand{\besseljs}[1]{\mathrm{J}_{#1}}
\newcommand{\besseljsp}[1]{\mathrm{J}_{#1}'}
\newcommand{\besseljspp}[1]{\mathrm{J}_{#1}''}
\newcommand{\hankel}[3]{\mathrm{H}_{#1}^{(#2)}(#3)}
\newcommand{\hankelp}[3]{\mathrm{H'}_{#1}^{(#2)}(#3)}
\newcommand{\hankels}[2]{\mathrm{H}_{#1}^{(#2)}}
\newcommand{\hankelsp}[2]{\mathrm{H}_{#1}'^{(#2)}}
\newcommand{\hankelspp}[2]{\mathrm{H}_{#1}''^{(#2)}}
\newcommand{\laplace}{\Delta}
\newcommand{\neff}{n_{\mathrm{eff}}}
\newcommand{\fexp}{f_{\mathrm{expt}}}
\newcommand{\ftheo}{f_{\mathrm{calc}}}
\newcommand{\nexp}{\tilde{n}}
\newcommand{\Gtheo}{\Gamma_{\mathrm{calc}}}
\newcommand{\Gexp}{\Gamma_{\mathrm{expt}}}
\newcommand{\Grad}{\Gamma_{\mathrm{rad}}}
\newcommand{\Gabs}{\Gamma_{\mathrm{abs}}}
\newcommand{\Gant}{\Gamma_{\mathrm{ant}}}
\newcommand{\rhof}{\rho_{\mathrm{fluc}}}
\newcommand{\rhofscl}{\rhof^{\mathrm{scl}}}
\newcommand{\rhofSS}{\rhof^{(\mathrm{ss})}}
\newcommand{\rhofnr}[1]{\rho_{#1 n_r}}
\newcommand{\rhow}{\rho_{\mathrm{Weyl}}}
\newcommand{\Nweyl}{N_{\mathrm{Weyl}}}
\newcommand{\rhot}{\hat{\rho}}
\newcommand{\rhotscl}{\rhot_{\mathrm{scl}}}
\newcommand{\rhotSS}{\rhot^{(\mathrm{ss})}}
\newcommand{\rhotnr}[1]{\tilde{\rho}_{#1 n_r}}
\newcommand{\fmin}{f_{\mathrm{min}}}
\newcommand{\kmin}{k_{\mathrm{min}}}
\newcommand{\fmax}{f_{\mathrm{max}}}
\newcommand{\kmax}{k_{\mathrm{max}}}
\newcommand{\fcrit}{f_{\mathrm{crit}}}
\newcommand{\kcrit}{k_{\mathrm{crit}}}
\newcommand{\po}{\mathrm{po}}
\newcommand{\lpo}{\ell_\po}
\newcommand{\lpeak}{\ell_\mathrm{peak}}
\newcommand{\lpeakscl}{\lpeak^\mathrm{scl}}
\newcommand{\lmax}{\ell_{\mathrm{max}}}
\newcommand{\alphacrit}{\alpha_{\mathrm{crit}}}
\newcommand{\alphainc}{\alpha_\mathrm{inc}}
\newcommand{\chico}{\chi_{\mathrm{co}}}
\newcommand{\Psiexp}{\Psi_\mathrm{expt}}
\newcommand{\Psitexp}{\tilde{\Psi}_\mathrm{expt}}
\newcommand{\Psimod}{\Psi_\mathrm{mod}}
\newcommand{\Psisup}{\Psi_\mathrm{sup}}
\newcommand{\reffig}[1]{\mbox{Fig.~\ref{#1}}}
\newcommand{\subreffig}[1]{\mbox{Fig. \subref{#1}}}
\newcommand{\refeq}[1]{\mbox{Eq.~(\ref{#1})}}
\newcommand{\refsec}[1]{\mbox{Sec.~\ref{#1}}}
\newcommand{\reftab}[1]{\mbox{Table \ref{#1}}}
\newcommand{\etal}{\textit{et al.\ }}
\newcommand{\dA}{A}
\newcommand{\dB}{B}
\newcommand{\FSR}{\mathrm{FSR}}
\newcommand{\dist}{D}
\newcommand{\depth}{l}
\renewcommand{\Re}[1]{\mathrm{Re}\left(#1\right)}
\renewcommand{\Im}[1]{\mathrm{Im}\left(#1\right)}

\hyphenation{re-so-nan-ce re-so-nan-ces ex-ci-ta-tion z-ex-ci-ta-tion di-elec-tric ap-pro-xi-ma-tion ra-dia-tion Me-cha-nics quan-tum pro-posed Con-cepts pro-duct Reh-feld ob-ser-va-ble Se-ve-ral rea-so-nable Ap-pa-rent-ly re-pe-ti-tions re-la-tive quan-tum su-per-con-duc-ting ap-pro-xi-mate cri-ti-cal mea-su-red}

\title{Experimental Observation of Localized Modes in a Dielectric Square Resonator}

\author{S. Bittner}
\affiliation{Institut f\"ur Kernphysik, Technische Universit\"at Darmstadt, D-64289 Darmstadt, Germany}
\affiliation{Laboratoire de Photonique Quantique et Mol{\'e}culaire, CNRS UMR 8537, Institut d'Alembert FR 3242, {\'E}cole Normale Sup{\'e}rieure de Cachan, F-94235 Cachan, France}
\author{E. Bogomolny}
\affiliation{Universit{\'e} Paris-Sud, CNRS, LPTMS, UMR 8626, Orsay, F-91405, France}
\author{B. Dietz}
\email{dietz@ikp.tu-darmstadt.de}
\author{M. Miski-Oglu}
\affiliation{Institut f\"ur Kernphysik, Technische Universit\"at Darmstadt, D-64289 Darmstadt, Germany}
\author{A. Richter}
\email{richter@ikp.tu-darmstadt.de}
\affiliation{Institut f\"ur Kernphysik, Technische Universit\"at Darmstadt, D-64289 Darmstadt, Germany}

\date{\today}

\begin{abstract}
We investigated the frequency spectra and field distributions of a dielectric square resonator in a microwave experiment. Since such systems cannot be treated analytically, the experimental studies of their properties are indispensable. The momentum representation of the measured field distributions shows that all resonant modes are localized on specific classical tori of the square billiard. Based on these observations a semiclassical model was developed. It shows excellent agreement with all but a single class of measured field distributions that will be treated separately.
\end{abstract}


\maketitle

\section{Introduction}
Dielectric microresonators are used for a wide range of applications, e.g., as microlasers, sensors, or building blocks for optical circuits \cite{Vahala2004, Matsko2005}. Most investigations focus on cavities with circular and deformed circular \cite{McCall1992} or polygonal shapes. The former can exhibit high quality factors or directional lasing emission \cite{Chern2003, Song2009a, Wang2009a}, whereas the latter are of interest, first, for applications like filters \cite{Li2006, Marchena2008} and, second, because the crystal structure of some materials naturally implies such a resonator geometry \cite{Vietze1998, Nobis2004, Wang2006}. In most experiments with microcavities in the infrared and optical regime only the resonance spectrum is accessible. This severely limits the possibilities of a detailed comparison with and verification of model calculations. Experiments with microwave resonators, on the other hand, can provide much richer and more detailed information including the field distributions inside the resonator over a large frequency range. Furthermore, microwave resonators can be manufactured with high precision and thus allow us to compare models with a defect-free, almost ideal experimental system. In the microwave experiment presented here, we concentrate on the simplest case of a polygonal cavity, a dielectric square resonator, which cannot be handled analytically even though the classical hard-wall square billiard is integrable. 

While the resonance frequencies and field distributions of dielectric resonators can be calculated via a variety of numerical methods, approximate models for specific resonators or types of modes are generally much easier to implement and to handle. At the same time they may provide more insight than is possible just from numerical solutions. A variety of models for the dielectric square resonator has been proposed \cite{Marcatili1969, Guo2003, Moon2003, Chern2004, Li2006, Lebental2007, Che2010a, Bittner2011a} that are based on similar assumptions, as, e.g., that the modes associated with the prominent resonances are localized along classical trajectories that impinge at the boundaries with an angle of incidence equal or close to $45^\circ$, like the diamond orbit. Similar results were obtained for pentagonal \cite{Lebental2007} and hexagonal \cite{Wiersig2003, Nobis2004, Li2006, Wang2006} resonators. However, there is experimental evidence that modes are also localized on other types of trajectories. This led to the development of refined models \cite{Poon2001, Fong2003, Huang2008, Yang2009a, Bittner2010}. Furthermore, the existing models were so far only applied to subsets of resonances and to cases with a specific refractive index, and never compared to measured field distributions. In order to investigate the exact nature of the modal localization in dielectric square resonators and to develop a comprehensive model, experiments with a macroscopic microwave resonator were performed and are reported in this article. Both the frequency spectrum and the near-field distributions in configuration and momentum space were investigated in the experiments. The latter evidenced that the resonant modes are localized on trajectories with specific momentum vectors. We will introduce a ray-based semiclassical model that is not restricted to specific groups of modes but indeed allows us to label all modes with quantum numbers and their symmetry class. It shows excellent agreement with the experimental findings for a wide range of the effective refractive index. Here we focus on the main characteristics of the model. A more detailed study will be presented in a subsequent publication \cite{BittnerPrep}.

\section{Experiment}
\begin{figure}[tb]
\begin{center}
\includegraphics[width = 7 cm]{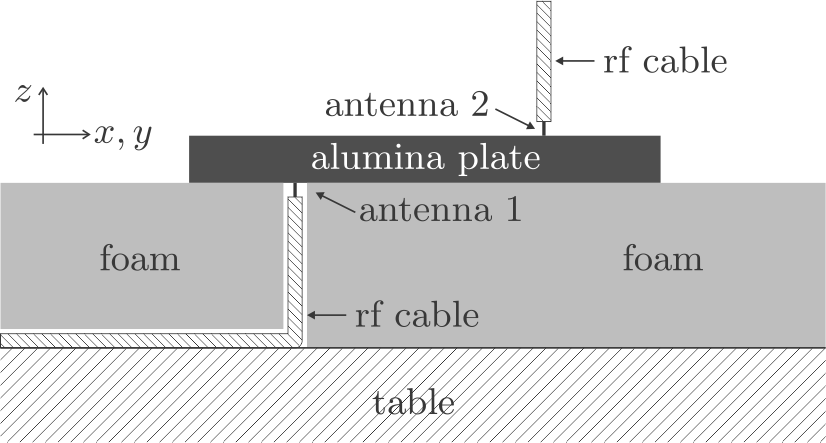}
\end{center}
\caption{Sectional drawing of the experimental setup (not to scale). The alumina plate is placed atop a foam with refractive index $\approx 1$. The two vertical wire antennas protruding from coaxial rf cables are connected to a VNA (adapted from Ref.~\cite{Bittner2011a}).}
\label{fig:expSetup}
\end{figure}

\begin{figure}[tb]
\begin{center}
\includegraphics[width = 8.4 cm]{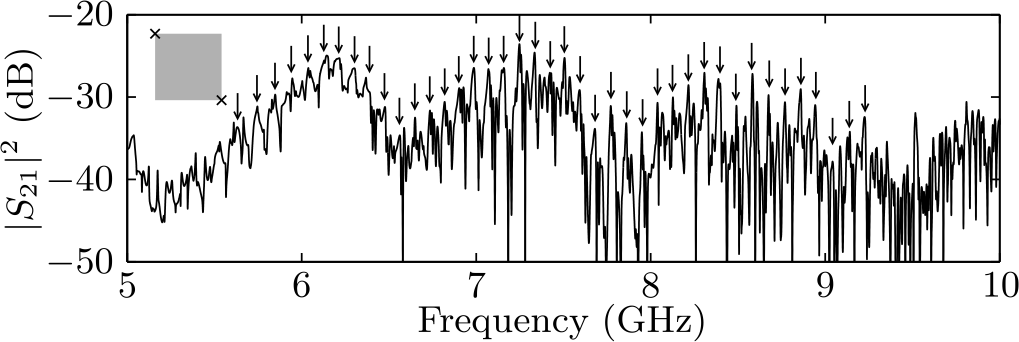}
\end{center}
\caption{Measured frequency spectrum in semilogarithmic scale. The arrows indicate a subset of roughly equidistant resonances. The inset indicates the positions of the antennas at the corners of the square resonator.}
\label{fig:spectrum}
\end{figure}

A sketch of the experimental setup is shown in \reffig{fig:expSetup}.
A ceramic plate made of alumina (Al$_2$O$_3$) with sharp corners and edges and refractive index $n_1 = 3.10$ was used as microwave resonator. The side length was $a = 297.30$~mm and the thickness was $b = 8.27$~mm, which was small compared to the wavelengths, $\lambda \approx 30$--$60$~mm, used in the experiments. Accordingly, the resonator is treated as a two-dimensional (2D) system by introducing the effective refractive index $\neff$ \cite{Bittner2009}, which is between $1.5$ and $2.5$ in the frequency range $f = 5.5$--$10.0$~GHz considered here. It was placed atop a $120$-mm-thick foam \cite{Rohacell} with a refractive index $n_2 = 1.02$ close to that of air to realize an effectively levitated resonator. Two wire antennas labeled $1$ and $2$ were positioned vertically to the resonator below and above it and connected to a vectorial network analyzer (VNA) \cite{PNA_N5230A}. The VNA measured the complex transmission amplitude $S_{21}(f)$ between the antennas, where $|S_{21}(f)|^2 = P_{2, \mathrm{out}} / P_{1, \mathrm{in}}$ is the ratio of the power coupled out of and into the resonator via antennas $2$ and $1$, respectively. An example of a measured frequency spectrum is shown in \reffig{fig:spectrum}.
The spectrum features a multitude of resonances with typical quality factors in the range of $Q = 200$--$2000$. Both transverse magnetic (TM) and transverse electric (TE) modes with the magnetic, respectively, electric field parallel to the plane of the resonator were excited. The polarization was determined using the procedure outlined in Ref.~\cite{Bittner2012a}. In the following, only the TM modes are considered since our setup is less sensitive to TE modes. In the frequency range $\approx 5.5$--$9.5$~GHz, a series of roughly equidistant resonances is observed as indicated by the arrows. In Refs.~\cite{Pan2003, Moon2003, Chern2004, Lebental2007, Huang2008, Bittner2010} these were generally associated with the diamond periodic orbit family. In addition, there are many other resonances yielding the overall complicated structure. The corresponding modes are localized on various families of classical trajectories as will be shown below. 

The field distributions inside the resonator were measured with the scanning antenna technique (cf.\ Refs.~\cite{Stein1995, Schaefer2006, Bittner2011}), i.e., the receiving antenna was moved around on the top surface of the resonator on a Cartesian grid with spatial resolution $\Delta_x = \Delta_y = a / 150 \approx 2$~mm. Here, the $(x, y) $plane is chosen parallel to that of the resonator and the $z$ direction perpendicular to it (cf.~\reffig{fig:expSetup}). The transmission amplitude is proportional to the $z$ component of the electric field vector, $E_z$, at the position $(x, y)$ of the antenna, so that for a TM mode the measurement of $S_{21}(x, y, f_\mathrm{res})$ at the resonance frequency $f_\mathrm{res}$ yields the corresponding field distribution $E_z(x, y)$, denoted as the measured wave function (WF) $\Psiexp(x, y)$ in the following.

\section{Analysis of Wave Functions}
\begin{figure*}[tb]
\begin{center}
\includegraphics[width = 13.5 cm]{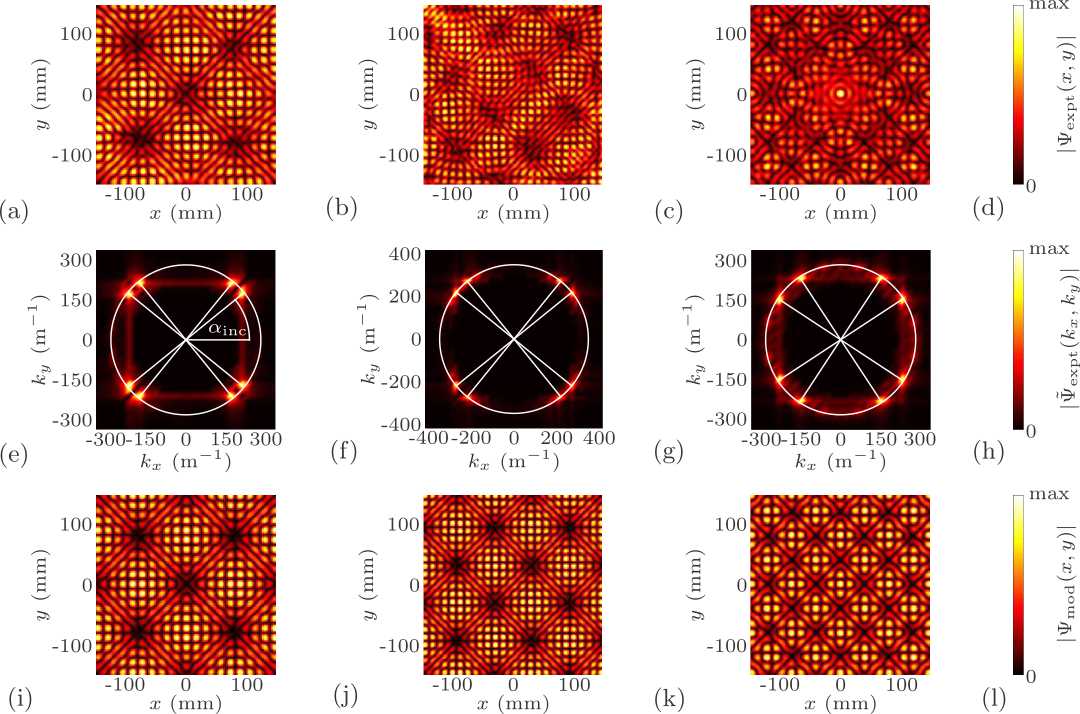}
\end{center}
\caption{(Color online) Examples of measured wave functions for (a) $6.835$~GHz, (b) $7.615$~GHz, and (c) $6.869$~GHz with co\-lor scale given in (d), and (e)--(g) the corresponding momentum distributions [with color scale in (h)]. They are identified with the model wave functions (i) TM$(16, 20, --)$ with $84.1\%$ overlap, (j) TM$(20,25,-+)$ with $67.1\%$ overlap, and (k) TM$(14,22,++)$ with $68.0\%$ overlap, respectively [color scale in (l)].}
\label{fig:WFexmpls}
\end{figure*}
Several measured WFs are shown in Figs.~\ref{fig:WFexmpls}(a)--\ref{fig:WFexmpls}(c). All are of a simple and clear structure reminiscent of that observed for integrable systems. Similar WFs were calculated numerically in Refs.~\cite{Lebental2007, Fong2003, Guo2003, Yang2009a, Che2010a}. Note that in some examples perturbations of the patterns around the position of the emitting antenna [e.g., in the center of \reffig{fig:WFexmpls}(c)] are observed. They originate from the direct transmission between the two antennas. 

\begin{figure}[tb]
\begin{center}
\includegraphics[width = 5.0 cm]{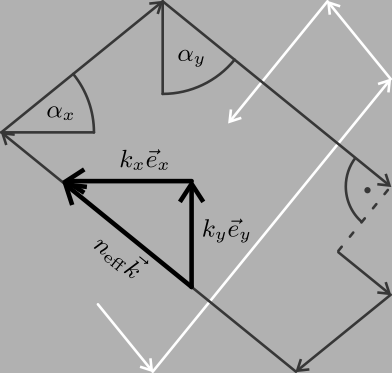}
\end{center}
\caption{Two examples of trajectories with the same angle of incidence (black and white thin arrows) belonging to a classical torus. The associated momentum vector $\neff \vec{k} = (k_x, k_y)$ (thick black arrows) corresponds to an angle of incidence of $\alpha_{x, y}$ on the vertical and horizontal boundaries of the square billiard, respectively.}
\label{fig:classTraj}
\end{figure}

A better understanding of the localization of WFs is achieved by considering their spatial Fourier transforms (FTs) \cite{Backer1999, Huang2002, Doya2007}, i.e., the corresponding momentum distributions $\Psitexp(k_x, k_y)$. Those corresponding to the WFs in Figs.~\ref{fig:WFexmpls}(a)--\ref{fig:WFexmpls}(c) are depicted in Figs.~\ref{fig:WFexmpls}(e)--\ref{fig:WFexmpls}(g). They exhibit a clear localization at eight specific momentum vectors $\neff \vec{k} = (k_x, k_y)$ pointing from the origin to the bright spots as indicated by the white lines. Such a localization is observed for all measured WFs. Note that in each panel the modulus of the momentum vectors, i.e., the radius of the white circles, equals $\neff k$, where $k = 2 \pi f / c$ with $c$ the speed of light in vacuum. Thus, the calculation of $\Psitexp(k_x, k_y)$ allows for the direct experimental determination of the effective refractive index. The eight momentum vectors correspond to a single family of classical trajectories that are specified by their angles of incidence $\alphainc$ [see \reffig{fig:WFexmpls}(e)] and $\pi/2 - \alphainc$, as shown in \reffig{fig:classTraj}.
Here, $\alphainc \in [0^\circ, 45^\circ]$ is defined as $\alphainc = \mathrm{min} \{ \alpha_x, \alpha_y \}$, and $\alpha_{x, y}$ are the angles of incidence on the vertical and horizontal boundaries of the square (cf.~\reffig{fig:classTraj}), respectively. In Figs.~\ref{fig:WFexmpls}(e)--\ref{fig:WFexmpls}(g) it has the values $\alphainc = 39.1^\circ$, $39.0^\circ$, and $33.3^\circ$, respectively. Note that only modes with $\alphainc \geq \alphacrit = \arcsin(1 / \neff)$ are observed experimentally.

\section{Ray-based Model}
\begin{table*}[tb]
\caption{Symmetry classes, quantum numbers and wave functions. The first column denotes the symmetry with respect to the diagonals, the second column is the symmetry with respect to both the horizontal and vertical axis, the third column is the parity of $m_x + m_y$, the fourth column the parity of $m_x$ and $m_y$ (which is the same for $++$ and $--$ modes but different for $+-$ and $-+$ modes), and the fifth column is the corresponding model wave function.}
\label{tab:dlmWFs}
\vspace{3 mm}
\begin{center}
\begin{tabular}{c|c|c|c|c}
\hline
\hline
Diag. & Horiz./vert. & Parity of & Parity of & Model wave function \\
sym. & sym. & $m_x + m_y$ & $m_x$, $m_y$ & \\
\hline
$++$ & $+$ & Even & Even & $\Psimod(x, y) = \cos(k_x x) \cos(k_y y) + \cos(k_y x) \cos(k_x y)$ \\
$++$ & $-$ & Even & Odd & $\Psimod(x, y) = \sin(k_x x) \sin(k_y y) + \sin(k_y x) \sin(k_x y)$ \\
$--$ & $+$ & Even & Even & $\Psimod(x, y) = \cos(k_x x) \cos(k_y y) - \cos(k_y x) \cos(k_x y)$ \\
$--$ & $-$ & Even & Odd & $\Psimod(x, y) = \sin(k_x x) \sin(k_y y) - \sin(k_y x) \sin(k_x y)$ \\
$+-$ & None & Odd & & $\Psimod(x, y) = \sin(k_x x) \cos(k_y y) + \cos(k_y x) \sin(k_x y)$ \\
$-+$ & None & Odd & & $\Psimod(x, y) = \sin(k_x x) \cos(k_y y) - \cos(k_y x) \sin(k_x y)$ \\
\hline
\hline
\end{tabular}
\end{center}
\end{table*}
A semiclassical model can be readily deduced on the basis of the dominant contributions from eight momentum vectors with angles of incidence determined by the classical reflection laws as illustrated in \reffig{fig:WFexmpls}. It is evident that such a phenomenon may be ascribed to an approximate quantization of classical tori (see \reffig{fig:classTraj}) in close analogy to quantum billiards with Dirichlet boundary conditions. In distinction to the latter, for an open dielectric square cavity the reflection coefficient is nontrivial after a reflection from the boundary. In the simplest approximation it is obtained from the Fresnel formulas, which are strictly speaking valid only for infinitely long interfaces, and the wave vector $\neff \vec{k} = (k_x, k_y)$ has to fulfill the conditions
\begin{equation} \label{eq:quantCond} \begin{array}{rcl} \exp\{ 2 i k_x a \} \, r^2(\alpha_x) & = & 1 \\ \exp\{ 2 i k_y a \} \, r^2(\alpha_y) & = & 1 \, . \end{array} \end{equation}
Here, $r(\alpha)$ denotes the Fresnel reflection coefficient for an angle of incidence $\alpha$ with respect to the surface normal,
\begin{equation} r(\alpha) = \frac{\neff \cos(\alpha) - \sqrt{1 - \neff^2 \sin^2(\alpha)}}{\neff \cos(\alpha) + \sqrt{1 - \neff^2 \sin^2(\alpha)}} \, , \end{equation}
and $\alpha_{x,y} = \arctan[\Re{k_{y, x}} / \Re{k_{x, y}}]$. The solutions of the coupled equations are calculated by iteration and can be formally written as
\begin{equation} \begin{array}{rcl} k_x & = & \{ \pi m_x + i \ln[r(\alpha_x)] \} / a \\ k_y & = & \{ \pi m_y + i \ln[r(\alpha_y)] \} / a \end{array} \end{equation}
with $m_{x, y} = 0, 1, 2, 3, \dots$ being the $x$ and $y$ quantum numbers \cite{Che2010a}. This yields for the associated resonance frequency $f_\mathrm{res} = c \sqrt{k_x^2 + k_y^2} / (2 \pi \neff)$. Models taking into account only the diamond periodic orbit \cite{Guo2003, Lebental2007, Bittner2011a} are recovered in the limit $\alphainc \approx 45^\circ$. The model WFs $\Psimod$ corresponding to our ansatz are superpositions of eight plane waves with momentum vectors $(\pm k_x, \pm k_y)$ and $(\pm k_y, \pm k_x)$. The associated amplitudes depend on the symmetry class of the mode. There are altogether six different symmetry classes in the square resonator \cite{McIsaac1975, Guo2003, Yang2007}. We label the modes by $(m_x, m_y, s_1 s_2)$ with $s_{1, 2} \in \{+, -\}$. Here $s_1 = +1$ ($s_2 = +1$) when $\Psimod$ is symmetric, and $s_1 = -1$ ($s_2 = -1$) when $\Psimod$ is antisymmetric with respect to the diagonal $x = y$ ($x = -y$). The model WFs $\Psimod$ and their symmetries with respect to the horizontal and vertical axes are listed in \reftab{tab:dlmWFs}. The $(m_x, m_y, -+)$ and $(m_x, m_y, +-)$ modes are degenerate due to symmetry reasons \cite{McIsaac1975}. The model furthermore predicts that this is also the case for the $(m_x, m_y, --)$ and $(m_x, m_y, ++)$ modes. In practice, however, they have slightly differing resonance frequencies due to their distinct features at the corners, where the $(--)$ modes have minimal intensity whereas the $(++)$ modes have maximal intensity. To identify the model WF that corresponds to a given measured WF $\Psiexp$, we calculated the overlap integral $|C|^2 = | \left< \Psiexp(f) \middle| \Psimod(m_x, m_y, s_1 s_2) \right> |^2$ for several trial functions $\Psimod$. For isolated resonances, a typical overlap of $|C|^2 = 50$--$80\%$ indicates that the corresponding model WF is the correct one, while the overlaps with model WFs not related to $\Psiexp$ are usually less than $3\%$. The specific value of an overlap depends on the data quality of the measured WFs and always takes a value below $100\%$. This is also the case for the numerically calculated WFs since the semiclassical model does not represent an exact solution of the Helmholtz equation. In some cases, the data quality of the measured WFs was too bad to allow for an identification. Three examples of model WFs are shown in the bottom row of \reffig{fig:WFexmpls}. The overlap of the WF $\Psimod(16, 20, --)$ in \reffig{fig:WFexmpls}(i) with the measured WF in \reffig{fig:WFexmpls}(a) is $84.1\%$, that of $\Psimod(20, 25, -+)$ in \reffig{fig:WFexmpls}(j) with the WF in \reffig{fig:WFexmpls}(b) is $67.1\%$, and that of $\Psimod(14, 22, ++)$ in \reffig{fig:WFexmpls}(k) with the WF in \reffig{fig:WFexmpls}(c) is $68.0\%$. Indeed, each measured WF with $m_x \neq m_y$ could be unambiguously identified with one model WF. Accordingly, the resonant modes for which clear WFs were obtained can be labeled by quantum numbers and allocated to a symmetry class as is the case for integrable systems. It turns out that the modes indicated by the arrows in \reffig{fig:spectrum} are those with the minimal possible $|m_x - m_y|$ for the given symmetry class, i.e., those that are localized closest to the diamond orbit with $\alphainc = 45^\circ$. The other resonances either correspond to TM modes with smaller values of $\alphainc$ or are TE polarized. It is expected that the quality factor decreases when the angle of localization $\alphainc$ deviates more and more from $45^\circ$. We, however, could not provide evidence of this behavior with our data since the contributions to the resonance widths are dominated by other mechanisms, such as the absorption in the alumina and the coupling losses due to the antennas. 

\begin{figure}[tb]
\begin{center}
\includegraphics[width = 7.5 cm]{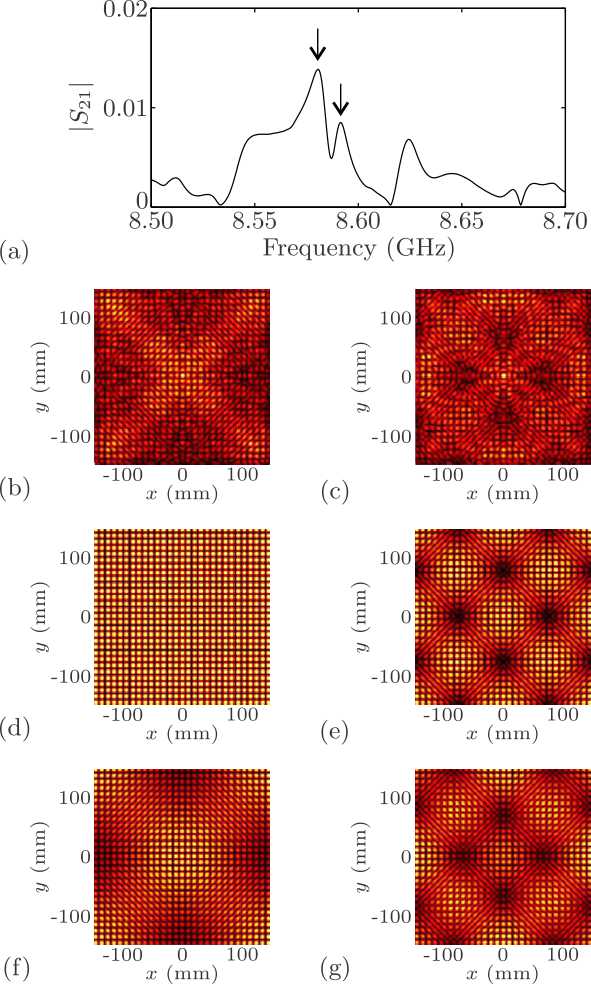}
\end{center}
\caption{(Color online) (a) Magnification of the frequency spectrum around $8.6$~GHz. (b), (c) Measured WFs for the resonances at $8.581$ and $8.591$~GHz indicated by  the left and right arrow in panel (a), respectively. (d) Model WF for TM$(28, 28, ++)$. (e) Model WF for TM$(26, 30, ++)$. (f), (g) Superpositions of model WFs corresponding to the measured WFs at $8.581$ and $8.591$~GHz, respectively.}
\label{fig:ppm}
\end{figure}

There is only one class of modes whose structure cannot be associated with a single model WF. An example, $\Psiexp(8.581~\mathrm{GHz})$, is shown in \reffig{fig:ppm}(b). The measured WF has an overlap of $62.4\%$ with $\Psimod(28, 28, ++)$, which is comparable to the values of the cases shown in \reffig{fig:WFexmpls}. Furthermore, its momentum distribution (not shown) exhibits a localization at $\alphainc = 45^\circ$, i.e., one along the trajectories from the family of the diamond orbit, like the model mode. However, the measured WF does not exhibit the chessboard structure observed for the model WF in \reffig{fig:ppm}(d).

 The frequency spectrum in \reffig{fig:ppm}(a) shows another resonance nearby at $8.591$~GHz. The two resonances are reasonably well isolated and exhibit only a small spectral overlap. The measured WF associated with the second resonance is shown in \reffig{fig:ppm}(c) and can be identified with the model mode TM$(26, 30, ++)$ [see \reffig{fig:ppm}(e)] with an overlap of $59.8\%$. Thus, the symmetry class associated with both resonances is the same. Its structure also shows deviations from that of the model WF. Such significant deviations from the patterns predicted by the model were not seen for any other classes of modes. Especially the chessboard pattern expected for the mode shown in \reffig{fig:ppm}(b) was not observed for any mode neither in the experiment nor in numerical calculations (not shown here). The overlaps between the measured WFs corresponding to the two resonances and the model WFs associated with the other one are, respectively, $|\left< \Psiexp(8.581~\mathrm{GHz}) \middle| \Psimod(26, 30, ++) \right>|^2 = 10.2\%$ and $|\left< \Psiexp(8.5891~\mathrm{GHz}) \middle| \Psimod(28, 28, ++) \right>|^2 = 11.1\%$. This cannot be attributed to the weak overlap of the two resonances in the frequency spectrum since for other pairs with similar features a significantly smaller overlap of the measured WFs with the model WFs corresponding to the neighboring resonance was observed. Thus, evidently there is a nonnegligible coupling between the two resonances that was also confirmed by numerical calculations. Consequently more than one model WF is needed to account for their structures. 
Indeed, the respective superpositions
\begin{equation} \begin{array}{l} \left| \Psisup(f) \right> = \\ \left< \Psimod(28, 28, ++) \middle| \Psiexp(f) \right> \left| \Psimod(28, 28, ++) \right> \\ + \left< \Psimod(26, 30, ++) \middle| \Psiexp(f) \right> \left| \Psimod(26, 30, ++) \right> \end{array} \end{equation}
agree well with the measured WFs [see Figs.~\ref{fig:ppm}(f) and \ref{fig:ppm}(g), respectively], as also confirmed by the corresponding overlaps $|\left< \Psiexp(8.581~\mathrm{GHz}) \middle| \Psisup(8.581~\mathrm{GHz}) \right>|^2 = 72.5\%$ and $|\left< \Psiexp(8.591~\mathrm{GHz}) \middle| \Psisup(8.591~\mathrm{GHz}) \right>|^2 = 70.9\%$. This situation is exemplary for all modes localized on $\alphainc = 45^\circ$, i.e., with $m_x = m_y$ and $++$ symmetry, and their neighboring modes [i.e., $(m_x-2, m_x+2, ++)$]. It should be noted that the $(m_x, m_x, ++)$ and the $(m_x-2, m_x+2, ++)$ model modes are closer in frequency than any other ones having the same symmetry class. This could be the reason why their mutual interaction becomes important. We suppose that these modes interact via a kind of tunneling effect. A detailed study of the effect, which we do not yet understand, is beyond the scope of the current article and will be reported in a future publication \cite{BittnerPrep}. A similar coupling effect for the modes with $m_x = m_y$ was observed in Ref.~\cite{Fong2003}, which was, however, attributed to the influence of the coupling to a waveguide, i.e., a perturbation of the square geometry.

\section{Conclusions}
We have measured the frequency spectrum and near-field distributions of a dielectric square resonator and demonstrated that the resonant states are localized close to classical tori of the square billiard that are quantized by taking into account the corresponding Fresnel reflection coefficients. We developed a simple but efficient semiclassical model that describes all the measured modes of the dielectric square resonator very well and allows us to label them unambiguously with quantum numbers. Note that to our knowledge this was possible hitherto only for dielectric resonators with circular shape \cite{Hentschel2002b}. Furthermore, we showed that the model works well for a large range of the effective refractive index ($\neff \approx 1.5$--$2.5$) and, in spite of its semiclassical nature, even in a frequency regime where the side length of the resonator is only five times larger than the free space wavelength, indicating that the range of its validity is not limited to the semiclassical regime. The accuracy of the model for even smaller size-to-wavelength ratios remains to be verified. Only for a single class of modes, i.e., those localized on trajectories with angle of incidence $45^\circ$, a small coupling between neighboring model modes of the same symmetry needs to be taken into account. Such a coupling effect has so far not been discussed for dielectric square resonators. The investigation of the underlying mechanism is of interest since ray-based models like the one proposed here are widely used and consequently an understanding of any wave-mechanical effects that limit their precision or validity is indispensable. A future project is the modeling of the far-field distributions, which are important for microlaser applications. Furthermore, the generalization of the model to other polygonal shapes like hexagons would be of great practical and theoretical interest.

\begin{acknowledgments}
\textit{Acknowledgments.}---
This work was supported by the DFG within the Sonderforschungsbereich 634.
\end{acknowledgments}

\end{document}